\documentclass[reprint,pre,amsmath,showpacs,superscriptaddress,bibnotes,twoside,eqsecnum]{revtex4-1}

\usepackage{amsfonts}
\usepackage{amssymb}
\usepackage{graphicx}
\usepackage{stmaryrd}
\usepackage{color}

\begin{document}

\title{Discrimination between two mechanisms of surface-scattering in a single-mode waveguide}

\author{M.~Rend\'{o}n}
\email{mrendon@ece.buap.mx}
\affiliation{Facultad de Ciencias de la Electr\'{o}nica, Universidad Aut\'{o}noma de Puebla, Puebla, Pue., 72570, M\'{e}xico}

\author{F.~M.~Izrailev}
\email{izrailev@sirio.ifuap.buap.mx}
\affiliation{Instituto de F\'{\i}sica, Universidad Aut\'{o}noma de Puebla, \\Apartado Postal J-48, Puebla, Pue., 72570, M\'{e}xico}

\author{N.~M.~Makarov}
\email{makarov@siu.buap.mx}
\affiliation{Instituto de Ciencias, Universidad Aut\'{o}noma de Puebla, \\Priv. 17 Norte No. 3417, Col. San Miguel Hueyotlipan, Puebla, Pue., 72050, M\'{e}xico}

\date{\today}

\begin{abstract}
Transport properties of a single-mode waveguide with rough boundary are studied by discrimination between two mechanisms of surface scattering, the \emph{amplitude} and \emph{square-gradient} ones. Although these mechanisms are generically mixed, we show that for some profiles they can separately operate within non-overlapping intervals of wave numbers of scattering waves. This effect may be important in realistic situations due to inevitable long-range correlations in scattering profiles.
\end{abstract}


\maketitle

\section{Introduction}
\label{sec:Intro}
The wave propagation along various guiding systems with rough boundaries is a typical situation in many fields of physics. The atmospheric and oceanic layers, as well as thin metal films and semiconductor nanostructures with growth defects and fractures, are the examples of such systems \cite{Chopra_book_1969, Knr_JAcoustSocAm_1974, BassFuks_book_1979, SntBrw_inbook_1986, TsnJrcMkw_PhysRevLett_1986, TrvAsh_PhysRevB_1988, BrtRsh_PhysRevB_1996, MyrStp_PhysRevB_1999, MyrStp_JPhysCM_2000, MyrPnm_PhysRevB_2002}. The rough waveguides with a prescribed surface profile can be prepared, for example, by the lithographic method in guiding electron/wave nanodevices \cite{WstODn_JOptSocAmA_1995, BllDzHyTnTrrPrr_PhysRevLett_1999}.

The propagating waves in rough guides are scattered by the inhomogeneities in the surface. We have already demonstrated that the \emph{surface scattering} consists of two basic mechanisms: the well-known \emph{amplitude} mechanism and the \emph{square-gradient} mechanism (see Refs.~\cite{IzrMkrRnd_PhysStatSolB_2005, IzrMkrRnd_PhysRevB_2005, IzrMkrRnd_PhysRevB_2006, RndIzrMkr_PhysRevB_2007}). Correspondingly, the inverse attenuation length is expressed via two terms. The term associated with the amplitude scattering depends on the variance of the disorder, $\sigma^2$, whereas the term associated with the square-gradient scattering is due to the squared variance, $\sigma^4$. In evaluating the contribution of each term solely by paying attention to the dependence on $\sigma$, one can arrive to a wrong conclusion that the square-gradient term is negligible compared to the amplitude one. It is important that the two terms differently depend on the roughness correlation parameters. For example, in the case of Gaussian correlations in the rough surface profile, these two mechanisms compete to each other, and the result of the competition essentially depends on the value of the wave number, see Refs.~\cite{IzrMkrRnd_PhysStatSolB_2005, IzrMkrRnd_PhysRevB_2005, IzrMkrRnd_PhysRevB_2006, RndIzrMkr_PhysRevB_2007, RndMkrIzr_PhysRevE_2011}.

On the other hand, it is known that the disordered waveguides with long-range correlations embedded in the rough surfaces, exhibit anomalous transport properties \cite{IzrMkr_OptLett_2001, IzrMak_PhysRevB_2003, IzrMkr_JPhysA_2005}. Specifically, with a proper design of the roughness profile one can strongly suppress the scattering in some interval of wave numbers, and, as a consequence, obtain the ballistic transport even for a very long waveguide. The goal of our paper is to study the conditions under which the amplitude and square-gradient mechanisms ``work" in different intervals of the wave numbers, and, therefore, do not compete to each other. In such a situation it is not correct to neglect any of these two mechanisms of scattering. We demonstrate that for some profiles one can observe the ``isolated" intervals of wave numbers where the attenuation of waves is due to only one of two mechanisms.

\section{The model}
\label{sec:Problem}

\begin{figure}[t!]
\includegraphics[width=8.6cm]{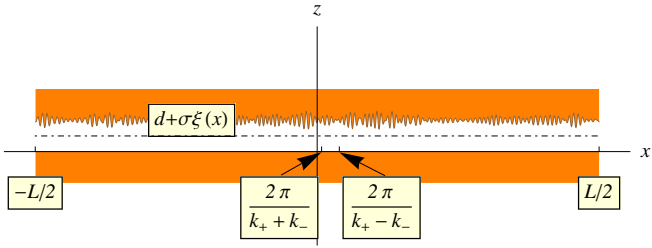}
\caption{\label{fig:SurfCorrDisorderedWaveguide}
(Color online) Waveguide with one rough boundary with correlated disorder of two characteristic scales $2\pi/(k_{+}+k_{-})$ and $2\pi/(k_{+}-k_{-})$.}
\end{figure}

We consider an open surface-disordered waveguide (or conducting wire) of length $L$ and average width $d$, stretched along the $x$- and $z$-axes, respectively, with $d\ll L$. The lower surface of the waveguide is taken, for simplicity, flat, $z=0$. The upper surface is supposed to be of a rough (corrugated) profile $z=d+\sigma \xi(x)$, slightly deviated from its flat average $z=d$. Here $\sigma$ is the root-mean-square \emph{roughness height}. In other words, the surface-corrugated guiding system occupies the area of the $(x,z)$-plane, defined by the relations (see Fig.~\ref{fig:SurfCorrDisorderedWaveguide})
\begin{equation}\label{eq:OccupiedRegion}
-L/2\leq x\leq L/2,\qquad 0\leq z\leq d+\sigma\xi(x).
\end{equation}
The surface corrugations are assumed to be small in height, $\sigma\ll d$. This limitation is common in the surface scattering theories that are based on appropriate perturbative approach, see, for example, Ref.~\onlinecite{BassFuks_book_1979}. The random function $\xi(x)$ describes the surface roughness and is assumed to be statistically homogeneous and isotropic, with the statistical properties of zero mean, $\langle\xi(x)\rangle=0$, unit variance, $\langle\xi^2(x)\rangle=1$, and binary correlation function
\begin{equation}\label{eq:<xixi>}
\langle\xi(x)\,\xi(x')\rangle={\cal W}(|x-x'|).
\end{equation}
The angular brackets represent the statistical averaging over different realizations of the random surface profile $\xi(x)$. The binary correlator ${\cal W}(x)$ has the normalization ${\cal W}(0)=1$.

As was predicted in Refs.~\onlinecite{IzrMkrRnd_PhysStatSolB_2005, IzrMkrRnd_PhysRevB_2005, IzrMkrRnd_PhysRevB_2006, RndIzrMkr_PhysRevB_2007}, the weak surface scattering in single-mode waveguides is determined by two different mechanisms: amplitude (A) and square-gradient (SG) surface scattering. Correspondingly, the transport is specified by two power spectra. The \emph{roughness-height} power spectrum is given by
\begin{equation}\label{eq:FT(W)}
W(k_x)=\int_{-\infty}^{\infty}dx\exp(-ik_xx)\,{\cal W}(x).
\end{equation}
The \emph{roughness-square-gradient} power spectrum is defined as
\begin{equation}\label{eq:FT(W''2)}
S(k_x)=\int_{-\infty}^{\infty}dx\exp{(-ik_xx)}\,{{\cal W}''}^2(x).
\end{equation}
The prime to the function denotes the derivative with respect to $x$. Since ${\cal W}(x)$ and ${\cal W}''^2(x)$ are real and even functions of $x$, their Fourier transforms are even and real functions of the longitudinal wave number $k_x$. It should be also stressed that according to rigorous mathematical theorem, the power spectra are non-negative functions of $k_x$ for any real random process $\xi(x)$.

We restrict our study to the single-mode waveguide for which the \emph{mode parameter}, $kd/\pi$, is confined to the relations
\begin{equation}\label{eq:1Mode_kRange}
1<kd/\pi<2.
\end{equation}
Here $k=\omega/c$ is the total wave number for the electromagnetic wave of frequency $\omega$ and TE polarization, propagating through a waveguide with perfectly conducting lateral walls. For the modeling of electron 1D structures, the wave number $k$ should be regarded as the Fermi wave number within the isotropic Fermi-liquid model. Because of the quantization of the transverse wave number $k_{z}=\pi/d$, the quantum value $k_{1}$ of the longitudinal wave number $k_x$ is given by
\begin{equation} \label{eq:k1}
k_{1}=\sqrt{k^2-(\pi/d)^2}.
\end{equation}
The transport through single-mode waveguides is provided by the lowest normal mode with number $n=1$ that propagates with real values of $k_1$. From Eq.~\eqref{eq:1Mode_kRange}, it follows that the wave number $k_1$ is confined within the interval
\begin{equation} \label{eq:1Mode_k1Range}
0<k_1 d/\pi<\sqrt{3}.
\end{equation}
All other waveguide modes with numbers $n\geq2$ having purely imaginary values of $k_x$, are evanescent and do not contribute to transport properties. Note that the weak surface-scattering condition $\sigma\ll d$ leads to the inequality $k_1\sigma\ll 1$.

\section{Disordered profile with long-range correlations}

\begin{figure}[t]
\includegraphics[width=7.8cm]{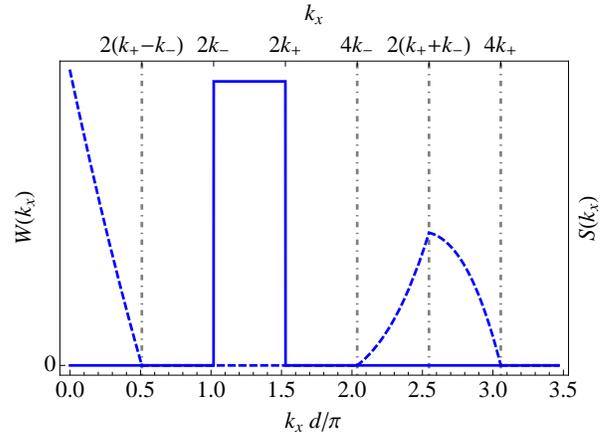}
\caption{\label{fig:PowerSpectra_CasoDeEstudio}
(Color online) Power spectra $W(k_x)$ (continuous line) and $S(k_x)$ (dashed line) under the condition \eqref{eq:condition2a_W_S}, for which the intervals where $W(k_x)\neq 0$ and $S(k_x)\neq 0$ do not overlap each other. Here $k_{-}d=1.6$ and $k_{+}d=2.4$, correspondingly, $k_{-}d/\pi=0.509$ and $k_{+}d/\pi=0.764$. The characteristic values of $k_x$ where singularities of the spectra occur, are shown at the top of the frame. The dimensionless variable $k_x d/\pi$ is indicated at the bottom of the frame.}
\end{figure}

In order to discriminate the amplitude scattering from the square-gradient one, we consider the correlator
\begin{equation} \label{eq:BinCorr1}
\mathcal{W}(x)=\frac{\sin(2k_{+}x)-\sin(2k_{-}x)}{2(k_{+}-k_{-})x},
\quad 0<k_{-}<k_{+},
\end{equation}
which gives rise, through Eq.~\eqref{eq:FT(W)}, to the power spectrum $W(k_x)$ of the following rectangular form (see continuous curve in Fig.~\ref{fig:PowerSpectra_CasoDeEstudio}),
\begin{equation} \label{eq:W1Spectrum}
W(k_x)=\frac{\pi}{2(k_{+}-k_{-})}\Theta(k_x-2k_{-})\,\Theta(2k_{+}-k_x).
\end{equation}
In Eqs.~\eqref{eq:BinCorr1} and \eqref{eq:W1Spectrum} the factor $1/2(k_+-k_-)$ provides the normalization requirement ${\cal W}(0)=1$. Here, the $\Theta(x)$ is the Heaviside unit-step function, $\Theta(x<0)=0$ and $\Theta(x>0)=1$. Recently, such a power spectrum became widely applicable in theoretical and experimental study of selective transport in 1D and quasi-1D disordered systems \cite{IzrMkr_JPhysA_2005,KhlIzrKrk_PhysRevLett_2008,DtzKhlStcMkrIzr_PhysRevB_2011}. The peculiarity of the random surfaces with such long-range correlations is that they have two characteristic scales, $2\pi/(k_{+}+k_{-})$ and $2\pi/(k_{+}-k_{-})$, see Fig.~\ref{fig:SurfCorrDisorderedWaveguide}. Consequently, the binary correlator and its power spectrum are specified by two correlation parameters.

With the substitution of Eq.~\eqref{eq:BinCorr1} into Eq.~\eqref{eq:FT(W''2)}, the following expression for the roughness-square-gradient power spectrum reads,
\begin{eqnarray} \label{eq:S1Spectrum}
&&S(k_x)=\frac{\pi}{4(k_{+}-k_{-})^2}\Big[\Theta(2 k_{+}-2 k_{-}-|k_x|)\,Q_1(k_x)\nonumber\\
&&+\Theta(|k_x|-4k_{-})\;\Theta(2k_{+}+2k_{-}-|k_x|)\,Q_2(k_x)\nonumber\\
&&+\Theta(|k_x| -2k_{+} -2k_{-})\;\Theta(4k_{+}-|k_x|)\,Q_3(k_x)\Big].\qquad
\end{eqnarray}
Here, the smooth functions $Q_1(k_x)$, $Q_2(k_x)$ and $Q_3(k_x)$ are described by
\begin{subequations}\label{eq:Qs}
\begin{eqnarray}
Q_1(k_x)&=&\frac{-|k_x|^5}{30}+\frac{8(k_{+}^3-k_{-}^3)k_x^2}{3}\nonumber\\
&-&8 (k{+}^4-k_{-}^4) |k_x|+\frac{32(k_{+}^5-k_{-}^5)}{5},\\
Q_2(k_x)&=&\frac{|k_x|^5}{60}-\frac{8 k_{-}^3 k_x^2}{3}+8k_{-}^4 |k_x|-\frac{32 k_{-}^5}{5},\\
Q_3(k_x)&=&\frac{-|k_x|^5}{60} +\frac{8 k_{+}^3 k_x^2}{3} -8 k_{+}^4 |k_x| +\frac{32 k_{+}^5}{5}.\qquad
\end{eqnarray}
\end{subequations}

The analysis of spectra \eqref{eq:W1Spectrum} and \eqref{eq:S1Spectrum} -- \eqref{eq:Qs} reveals the following. At $k_x>0$, $S(k_x)$ has two intervals of finite values, $0<k_x<2k_{+}-2k_{-}$ and $4k_{-}<k_x<4k_{+}$. When $k_{+}/2\leq k_{-}$ (with $k_{-}<k_{+}$, by definition), it follows that
\begin{equation} \label{eq:condition2a_W_S}
(k_{+}-k_{-})\leq k_{-}<k_{+}\leq 2k_{-}.
\end{equation}
Then, the above intervals where $S(k_x)\neq 0$, are well spaced for not to overlap with the interval $2k_{-}<k_x<2k_{+}$, where $W(k_x)\neq 0$. The condition \eqref{eq:condition2a_W_S} is illustrated in Fig.~\ref{fig:PowerSpectra_CasoDeEstudio} (with $S(k_x)$ represented by the dashed line). This case is of great importance since the amplitude and square-gradient scattering emerge alone in different intervals of the wave number.

In accordance with the convolution method, discussed in detail in, e.g., Refs.~\onlinecite{IzrMkr_JPhysA_2005,TsnKngDng_book_2001}, the random surface profile $\xi(x)$ with the properties \eqref{eq:BinCorr1}, \eqref{eq:W1Spectrum} and \eqref{eq:S1Spectrum} -- \eqref{eq:Qs}, can be constructed with the use of the following expression:
\begin{equation}\label{xi-WOD}
\xi(x)=\int_{-\infty}^\infty\,\frac{dx'}{\sqrt{2\pi}}\,Z(x-x')\,
\frac{\sin(2k_+x')-\sin(2k_-x')}{(k_+-k_-)^{1/2}x'}.
\end{equation}
Here, the white noise $Z(x)$ is defined by the standard properties, $\langle Z(x)\rangle=0$, $\langle Z(x) Z(x')\rangle=\delta(x-x')$. Its discrete version can be easily created by random number generators. Figure~\ref{fig:SurfCorrDisorderedWaveguide} displays one of the realizations of such a surface profile.

\section{Transmittance and localization length}

As is known (see, e.g., Ref.~\onlinecite{IzrMkr_JPhysA_2005}), the transport properties of a single-mode weakly disordered waveguide is governed by the phenomenon of Anderson localization. Therefore, its average transmittance $\langle T\rangle$ can be described by the expression
\begin{equation}\label{eq:<T>}
\begin{split}
\langle T(L/L_{\text{loc}})\rangle=&\frac{1}{2\sqrt{\pi}}\left(\frac{L}{2L_{\text{loc}}}
\right)^{-3/2}\exp\left(-\frac{L}{2L_{\text{loc}}}\right)\\
&\times\int_0^\infty\frac{z^2\,dz}{\cosh z}\exp\left(-z^2\frac{L_{\text{loc}}}{2L}\right).
\end{split}
\end{equation}
This function depends solely on the single scaling parameter $L/L_{\text{loc}}$ with $L_{\text{loc}}$ being the so-called localization length. The localization length $L_{\text{loc}}$ is associate with the backscattering length $L_{\text{bs}}$ that can be readily extracted from the corresponding expression for the mode-attenuation length calculated in Ref.~\onlinecite{IzrMkrRnd_PhysRevB_2006}. Taking into account the relation $2L_{\text{bs}}=L_{\text{loc}}$, which is specific for Ref.~\onlinecite{IzrMkrRnd_PhysRevB_2006}, one can reveal that when the number of conducting channels in the smooth waveguide equals to one, the inverse localization length of the first propagating mode is given by
\begin{subequations} \label{eq:L(AG)(SG)(b)}
\begin{eqnarray}
\label{eq:L(b)}
\frac{1}{L_{\text{loc}}}&=&\frac{1}{L_{\text{loc}}^{(A)}}+\frac{1}{L_{\text{loc}}^{(SG)}}, \\
\label{eq:L(AG)(b)}
\frac{1}{L_{\text{loc}}^{(A)}}&=&\frac{\sigma^2}{d^6} \frac{\pi^4}{2k_1^2} W(2k_1), \\
\label{eq:L(SG)(b)}
\frac{1}{L_{\text{loc}}^{(SG)}}&=&\frac{\sigma^4}{d^4} \frac{(3+4\pi^2)^2}{576\,k_1^2}
S(2k_1).
\end{eqnarray}
\end{subequations}
Here the first term is contributed by the conventional amplitude surface-scattering while the second one emerges exclusively due to the square-gradient surface-scattering. We stress that the term \eqref{eq:L(SG)(b)} was typically neglected in the literature since it is proportional to the squared variance of disorder, $\sigma^4$, and seems to be much smaller than the term \eqref{eq:L(AG)(b)} with the $\sigma^2$-dependence. However, apart from the disorder strength, one has to take into account the correlation properties of scattering profiles. The square-gradient scattering mechanism is much more sophisticated and sensitive to specific correlations, than the relatively simple amplitude surface-scattering mechanism.

\section{Discriminated surface scattering mechanisms}

\begin{figure}
\includegraphics[width=7.7cm]{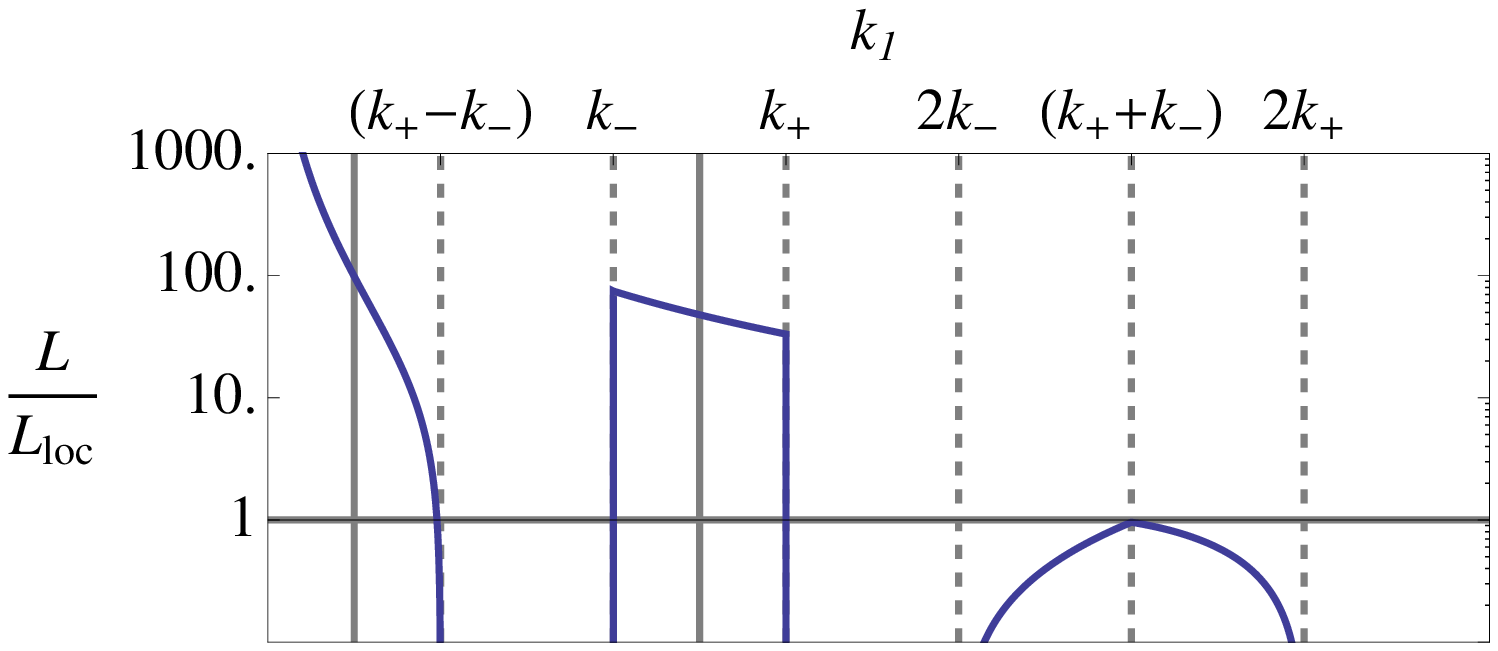} \\
\hspace{0.02cm} \includegraphics[width=7.55cm]{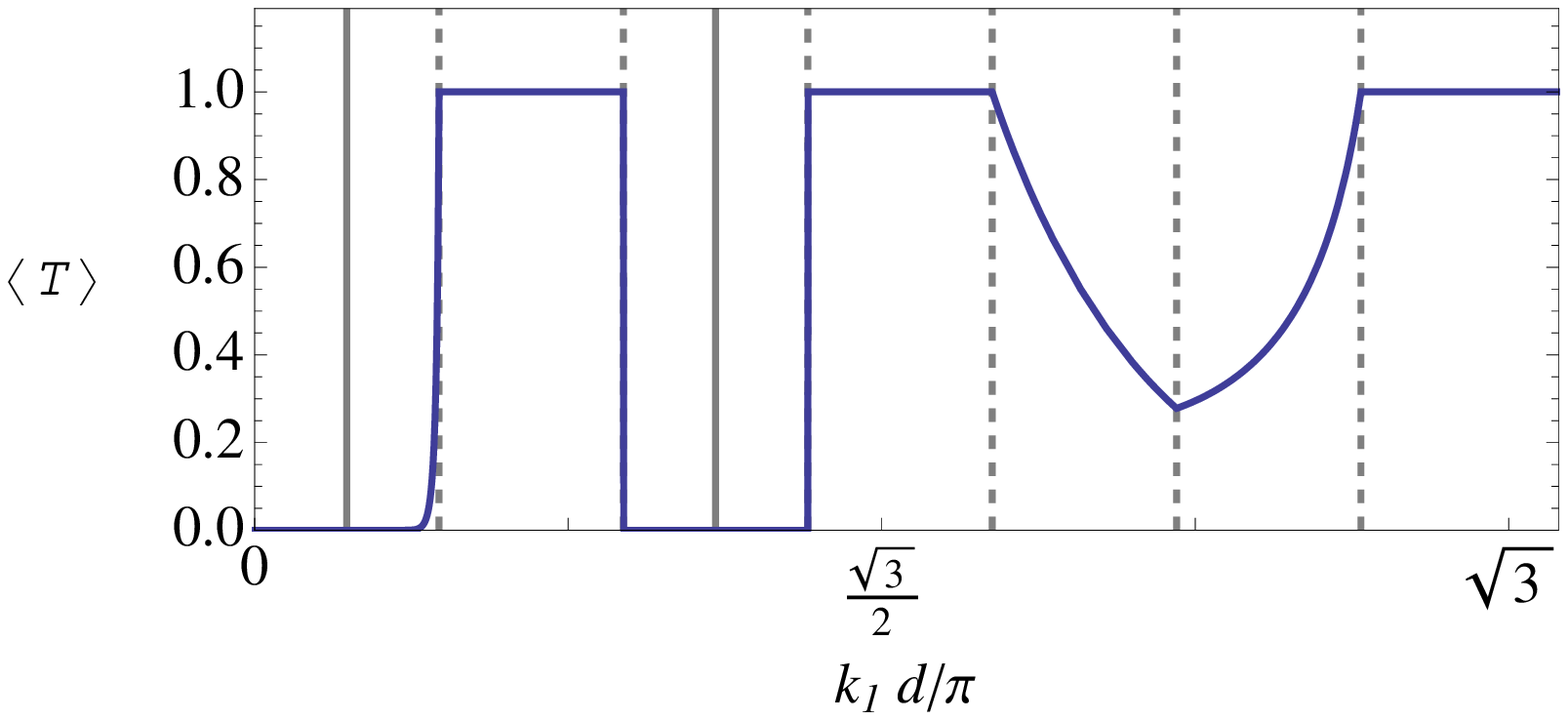}
\caption{\label{fig:plots1RBVsk1dOverPi}
(Color online) Plots of $L/L_{\text{loc}}$ and $\langle T\rangle$ vs. $k_1d/\pi$ for the parameters
$k_{-}d=1.6$, $k_{+}d=2.4$ ($k_{-}d/\pi=0.509$, $k_{+}d/\pi=0.764$),
$\sigma/d=1/10$ and $L/d=200$. Vertical solid lines mark the
values $k_1=(k_{+}-k_{-})/2$ and $k_1=(k_{+}+k_{-})/2$.}
\end{figure}

In order to recognize the square-gradient surface-scattering and clearly discriminate it from the conventional (amplitude) one, we assume the condition \eqref{eq:condition2a_W_S} to be met. The ratio $L/L_{\text{loc}}$, and the average transmittance $\langle T(L/L_{\text{loc}})\rangle$ are depicted in Fig.~\ref{fig:plots1RBVsk1dOverPi}. The localization effects arise within three separated intervals. In the first and third intervals, $0<k_1<(k_{+}-k_{-})$ and $2k_{-}<k_1<2k_{+}$, we have $1/L_{\text{loc}}^{(A)}=0$, and only the square-gradient scattering decrements the transmittance. In the first one the transmittance $\langle T\rangle=\langle T(L/L_{\text{loc}}^{(SG)})\rangle$ is almost zero since $L_{\text{loc}}^{(SG)}\ll L$. In the third interval $L_{\text{loc}}^{(SG)}\approx L$, and the transmittance $\langle T(L/L_{\text{loc}}^{(SG)})\rangle$ decreases up to the value $0.27$. Otherwise, within the second interval, $k_{-}<k_1<k_{+}$, only the amplitude scattering arises because $1/L_{\text{loc}}^{(SG)}=0$. Here, $\langle T\rangle=\langle T(L/L_{\text{loc}}^{(A)})\rangle$ is exponentially small since $L_{\text{loc}}^{(A)}\ll L$. The above three intervals are well separated by windows, $(k_{+}-k_{-})<k_1<k_{-}$ and $k_{+}<k_1<2k_{-}$, of ballistic transport where $\langle T\rangle\approx1$ due to divergence of both localization lengths. Naturally, for $2k_{+}<k_1$ the ballistic transport emerges again.

Because different scattering mechanisms manifest themselves in different intervals of $k_1$, the relation between partial localization lengths $L_{\text{loc}}^{(A)}$ and $L_{\text{loc}}^{SG)}$ does not crucial matter. Nevertheless, in order to complete the analysis, we write down the ratio of those quantities at two characteristic points (see the indication of these points as vertical solid lines in Fig.~\ref{fig:plots1RBVsk1dOverPi}),
\begin{equation} \label{eq:ratioApprox}
\frac{L_{\text{loc}}^{(A)}|_{k_1=(k_{+}+k_{-})/2}}{L_{\text{loc}}^{(SG)}|_{k_1=(k_{+}-k_{-})/2}}
\approx (\sigma/d)^2\, (k_{-} d)^4\; F(k_{+}/k_{-}).
\end{equation}
Here, the function of $F(\kappa)$ with $1<\kappa \leq 2$, reads
\begin{equation} \label{eq:F}
\begin{split}
F(\kappa)\approx&\frac{1.072\times 10^{-3}\;(\kappa+1)^2}{(\kappa-1)^2}(31\,\kappa^4+116\,\kappa^3\\
&+186\,\kappa^2+116\,\kappa+31)
\end{split}
\end{equation}
This function reaches its minimum $F(2)\approx24.1$ at $\kappa\approx2$ and monotonically increases when $\kappa$ decreases to $\kappa=1$, where it diverges. In Fig.~\ref{fig:plots1RBVsk1dOverPi} with $\kappa=1.5$ and $F(1.5)\approx31.4$, the ratio \eqref{eq:ratioApprox} can be approximated to $2.1$.\\

\section{Conclusions}

In this paper we have demonstrated that two different mechanisms of surface scattering, namely, the amplitude and square-gradient ones, can be discriminated in waveguides with correlated profiles. By discrimination here we mean that these two mechanisms are responsible for the scattering in a quite peculiar way, when the first type of scattering occurs in its own interval of wave number, and the second one emerges in other intervals. In such a situation, they are {\it both} important and neither one or another can be neglected. This effect is entirely due to a specific dependence of the two mechanisms on the correlation properties of scattering profiles. The example we have studied above is important for understanding the nature of the two types of surface scattering. We have to stress that the situation in which such an effect can arise is not only of the academic interest, and can emerge in real devices.

\begin{acknowledgments}
This work was supported by the SEP-CONACYT (M\'exico) grant No~166382.
\end{acknowledgments}

\bibliography{Q1D-RoughWaveguide_02}

\begin{thebibliography}{23}%
\makeatletter
\providecommand \@ifxundefined [1]{%
 \@ifx{#1\undefined}
}%
\providecommand \@ifnum [1]{%
 \ifnum #1\expandafter \@firstoftwo
 \else \expandafter \@secondoftwo
 \fi
}%
\providecommand \@ifx [1]{%
 \ifx #1\expandafter \@firstoftwo
 \else \expandafter \@secondoftwo
 \fi
}%
\providecommand \natexlab [1]{#1}%
\providecommand \enquote  [1]{``#1''}%
\providecommand \bibnamefont  [1]{#1}%
\providecommand \bibfnamefont [1]{#1}%
\providecommand \citenamefont [1]{#1}%
\providecommand \href@noop [0]{\@secondoftwo}%
\providecommand \href [0]{\begingroup \@sanitize@url \@href}%
\providecommand \@href[1]{\@@startlink{#1}\@@href}%
\providecommand \@@href[1]{\endgroup#1\@@endlink}%
\providecommand \@sanitize@url [0]{\catcode `\\12\catcode `\$12\catcode
  `\&12\catcode `\#12\catcode `\^12\catcode `\_12\catcode `\%12\relax}%
\providecommand \@@startlink[1]{}%
\providecommand \@@endlink[0]{}%
\providecommand \url  [0]{\begingroup\@sanitize@url \@url }%
\providecommand \@url [1]{\endgroup\@href {#1}{\urlprefix }}%
\providecommand \urlprefix  [0]{URL }%
\providecommand \Eprint [0]{\href }%
\@ifxundefined \urlstyle {%
  \providecommand \doi  [0]{\begingroup \@sanitize@url \@doi}%
  \providecommand \@doi [1]{\endgroup \@@startlink {\doibase
  #1}doi:\discretionary {}{}{}#1\@@endlink }%
}{%
  \providecommand \doi  [0]{doi:\discretionary{}{}{}\begingroup
  \urlstyle{rm}\Url }%
}%
\providecommand \doibase [0]{http://dx.doi.org/}%
\providecommand \Doi [0]{\begingroup \@sanitize@url \@Doi }%
\providecommand \@Doi  [1]{\endgroup\@@startlink{\doibase#1}\@@Doi}%
\providecommand \@@Doi [1]{#1\@@endlink}%
\providecommand \selectlanguage [0]{\@gobble}%
\providecommand \bibinfo  [0]{\@secondoftwo}%
\providecommand \bibfield  [0]{\@secondoftwo}%
\providecommand \translation [1]{[#1]}%
\providecommand \BibitemOpen [0]{}%
\providecommand \bibitemStop [0]{}%
\providecommand \bibitemNoStop [0]{.\EOS\space}%
\providecommand \EOS [0]{\spacefactor3000\relax}%
\providecommand \BibitemShut  [1]{\csname bibitem#1\endcsname}%
\bibitem [{\citenamefont {Chopra}(1969)}]{Chopra_book_1969}%
  \BibitemOpen
  \bibfield  {author} {\bibinfo {author} {\bibfnamefont {K.~L.}\ \bibnamefont
  {Chopra}},\ }\href@noop {} {\emph {\bibinfo {title} {Thin film phenomena}}}\
  (\bibinfo  {publisher} {McGraw-Hill},\ \bibinfo {address} {New York},\
  \bibinfo {year} {1969})\BibitemShut {NoStop}%
\bibitem [{\citenamefont {Konrady}(1974)}]{Knr_JAcoustSocAm_1974}%
  \BibitemOpen
  \bibfield  {author} {\bibinfo {author} {\bibfnamefont {J.~A.}\ \bibnamefont
  {Konrady}},\ }\href@noop {} {\bibfield  {journal} {\bibinfo  {journal} {J.
  Acoust. Soc. Am.},\ }\textbf {\bibinfo {volume} {56}},\ \bibinfo {pages}
  {1687} (\bibinfo {year} {1974})}\BibitemShut {NoStop}%
\bibitem [{\citenamefont {Bass}\ and\ \citenamefont
  {Fuks}(1979)}]{BassFuks_book_1979}%
  \BibitemOpen
  \bibfield  {author} {\bibinfo {author} {\bibfnamefont {F.~G.}\ \bibnamefont
  {Bass}}\ and\ \bibinfo {author} {\bibfnamefont {I.~M.}\ \bibnamefont
  {Fuks}},\ }\href@noop {} {\emph {\bibinfo {title} {Wave Scattering from
  Statistically Rough Surfaces}}}\ (\bibinfo  {publisher} {Pergamon},\ \bibinfo
  {address} {New York},\ \bibinfo {year} {1979})\BibitemShut {NoStop}%
\bibitem [{\citenamefont {DeSanto}\ and\ \citenamefont
  {Brown}(1986)}]{SntBrw_inbook_1986}%
  \BibitemOpen
  \bibfield  {author} {\bibinfo {author} {\bibfnamefont {J.~A.}\ \bibnamefont
  {DeSanto}}\ and\ \bibinfo {author} {\bibfnamefont {G.~S.}\ \bibnamefont
  {Brown}},\ }\enquote {\bibinfo {title} {Analytical techniques for multiple
  scattering from rough surfaces},}\ \ (\bibinfo  {publisher} {Elsevier},\
  \bibinfo {address} {Amsterdam},\ \bibinfo {year} {1986})\ Chap.~\bibinfo
  {chapter} {1}, pp.\ \bibinfo {pages} {1--62},\ \bibinfo {note} {in
  \emph{Progress in Optics}, Vol. 23, edited by E. Wolf}\BibitemShut {NoStop}%
\bibitem [{\citenamefont {Te\ifmmode \check{s}\else
  \v{s}\fi{}anovi\ifmmode~\acute{c}\else \'{c}\fi{}}\ \emph
  {et~al.}(1986)\citenamefont {Te\ifmmode \check{s}\else
  \v{s}\fi{}anovi\ifmmode~\acute{c}\else \'{c}\fi{}}, \citenamefont
  {Jari\ifmmode~\acute{c}\else \'{c}\fi{}},\ and\ \citenamefont
  {Maekawa}}]{TsnJrcMkw_PhysRevLett_1986}%
  \BibitemOpen
  \bibfield  {author} {\bibinfo {author} {\bibfnamefont {Z.}~\bibnamefont
  {Te\ifmmode \check{s}\else \v{s}\fi{}anovi\ifmmode~\acute{c}\else
  \'{c}\fi{}}}, \bibinfo {author} {\bibfnamefont {M.~V.}\ \bibnamefont
  {Jari\ifmmode~\acute{c}\else \'{c}\fi{}}}, \ and\ \bibinfo {author}
  {\bibfnamefont {S.}~\bibnamefont {Maekawa}},\ }\Doi
  {10.1103/PhysRevLett.57.2760} {\bibfield  {journal} {\bibinfo  {journal}
  {Phys. Rev. Lett.},\ }\textbf {\bibinfo {volume} {57}},\ \bibinfo {pages}
  {2760} (\bibinfo {year} {1986})}\BibitemShut {NoStop}%
\bibitem [{\citenamefont {Trivedi}\ and\ \citenamefont
  {Ashcroft}(1988)}]{TrvAsh_PhysRevB_1988}%
  \BibitemOpen
  \bibfield  {author} {\bibinfo {author} {\bibfnamefont {N.}~\bibnamefont
  {Trivedi}}\ and\ \bibinfo {author} {\bibfnamefont {N.~W.}\ \bibnamefont
  {Ashcroft}},\ }\Doi {10.1103/PhysRevB.38.12298} {\bibfield  {journal}
  {\bibinfo  {journal} {Phys. Rev. B},\ }\textbf {\bibinfo {volume} {38}},\
  \bibinfo {pages} {12298} (\bibinfo {year} {1988})}\BibitemShut {NoStop}%
\bibitem [{\citenamefont {Bratkovsky}\ and\ \citenamefont
  {Rashkeev}(1996)}]{BrtRsh_PhysRevB_1996}%
  \BibitemOpen
  \bibfield  {author} {\bibinfo {author} {\bibfnamefont {A.~M.}\ \bibnamefont
  {Bratkovsky}}\ and\ \bibinfo {author} {\bibfnamefont {S.~N.}\ \bibnamefont
  {Rashkeev}},\ }\Doi {10.1103/PhysRevB.53.13074} {\bibfield  {journal}
  {\bibinfo  {journal} {Phys. Rev. B},\ }\textbf {\bibinfo {volume} {53}},\
  \bibinfo {pages} {13074} (\bibinfo {year} {1996})}\BibitemShut {NoStop}%
\bibitem [{\citenamefont {Meyerovich}\ and\ \citenamefont
  {Stepaniants}(1999)}]{MyrStp_PhysRevB_1999}%
  \BibitemOpen
  \bibfield  {author} {\bibinfo {author} {\bibfnamefont {A.~E.}\ \bibnamefont
  {Meyerovich}}\ and\ \bibinfo {author} {\bibfnamefont {A.}~\bibnamefont
  {Stepaniants}},\ }\Doi {10.1103/PhysRevB.60.9129} {\bibfield  {journal}
  {\bibinfo  {journal} {Phys. Rev. B},\ }\textbf {\bibinfo {volume} {60}},\
  \bibinfo {pages} {9129} (\bibinfo {year} {1999})}\BibitemShut {NoStop}%
\bibitem [{\citenamefont {Meyerovich}\ and\ \citenamefont
  {Stepaniants}(2000)}]{MyrStp_JPhysCM_2000}%
  \BibitemOpen
  \bibfield  {author} {\bibinfo {author} {\bibfnamefont {A.~E.}\ \bibnamefont
  {Meyerovich}}\ and\ \bibinfo {author} {\bibfnamefont {A.}~\bibnamefont
  {Stepaniants}},\ }\href@noop {} {\bibfield  {journal} {\bibinfo  {journal}
  {J. Phys.: Condens. Matter},\ }\textbf {\bibinfo {volume} {12}},\ \bibinfo
  {pages} {5575} (\bibinfo {year} {2000})}\BibitemShut {NoStop}%
\bibitem [{\citenamefont {Meyerovich}\ and\ \citenamefont
  {Ponomarev}(2002)}]{MyrPnm_PhysRevB_2002}%
  \BibitemOpen
  \bibfield  {author} {\bibinfo {author} {\bibfnamefont {A.~E.}\ \bibnamefont
  {Meyerovich}}\ and\ \bibinfo {author} {\bibfnamefont {I.~V.}\ \bibnamefont
  {Ponomarev}},\ }\Doi {10.1103/PhysRevB.65.155413} {\bibfield  {journal}
  {\bibinfo  {journal} {Phys. Rev. B},\ }\textbf {\bibinfo {volume} {65}},\
  \bibinfo {pages} {155413} (\bibinfo {year} {2002})}\BibitemShut {NoStop}%
\bibitem [{\citenamefont {West}\ and\ \citenamefont
  {O'Donnell}(1995)}]{WstODn_JOptSocAmA_1995}%
  \BibitemOpen
  \bibfield  {author} {\bibinfo {author} {\bibfnamefont {C.~S.}\ \bibnamefont
  {West}}\ and\ \bibinfo {author} {\bibfnamefont {K.~A.}\ \bibnamefont
  {O'Donnell}},\ }\Doi {10.1364/JOSAA.12.000390} {\bibfield  {journal}
  {\bibinfo  {journal} {J. Opt. Soc. Am. A},\ }\textbf {\bibinfo {volume}
  {12}},\ \bibinfo {pages} {390} (\bibinfo {year} {1995})}\BibitemShut
  {NoStop}%
\bibitem [{\citenamefont {Bellani}\ \emph {et~al.}(1999)\citenamefont
  {Bellani}, \citenamefont {Diez}, \citenamefont {Hey}, \citenamefont {Toni},
  \citenamefont {Tarricone}, \citenamefont {Parravicini}, \citenamefont
  {Dom\'\i{}nguez-Adame},\ and\ \citenamefont
  {G\'omez-Alcal\'a}}]{BllDzHyTnTrrPrr_PhysRevLett_1999}%
  \BibitemOpen
  \bibfield  {author} {\bibinfo {author} {\bibfnamefont {V.}~\bibnamefont
  {Bellani}}, \bibinfo {author} {\bibfnamefont {E.}~\bibnamefont {Diez}},
  \bibinfo {author} {\bibfnamefont {R.}~\bibnamefont {Hey}}, \bibinfo {author}
  {\bibfnamefont {L.}~\bibnamefont {Toni}}, \bibinfo {author} {\bibfnamefont
  {L.}~\bibnamefont {Tarricone}}, \bibinfo {author} {\bibfnamefont {G.~B.}\
  \bibnamefont {Parravicini}}, \bibinfo {author} {\bibfnamefont
  {F.}~\bibnamefont {Dom\'\i{}nguez-Adame}}, \ and\ \bibinfo {author}
  {\bibfnamefont {R.}~\bibnamefont {G\'omez-Alcal\'a}},\ }\Doi
  {10.1103/PhysRevLett.82.2159} {\bibfield  {journal} {\bibinfo  {journal}
  {Phys. Rev. Lett.},\ }\textbf {\bibinfo {volume} {82}},\ \bibinfo {pages}
  {2159} (\bibinfo {year} {1999})}\BibitemShut {NoStop}%
\bibitem [{\citenamefont {Izrailev}\ \emph
  {et~al.}(2005){\natexlab{a}}\citenamefont {Izrailev}, \citenamefont
  {Makarov},\ and\ \citenamefont {Rend\'on}}]{IzrMkrRnd_PhysStatSolB_2005}%
  \BibitemOpen
  \bibfield  {author} {\bibinfo {author} {\bibfnamefont {F.~M.}\ \bibnamefont
  {Izrailev}}, \bibinfo {author} {\bibfnamefont {N.~M.}\ \bibnamefont
  {Makarov}}, \ and\ \bibinfo {author} {\bibfnamefont {M.}~\bibnamefont
  {Rend\'on}},\ }\Doi {10.1002/pssb.200460769} {\bibfield  {journal} {\bibinfo
  {journal} {phys. stat. sol. (b)},\ }\textbf {\bibinfo {volume} {242}},\
  \bibinfo {pages} {1224} (\bibinfo {year} {2005}{\natexlab{a}})}\BibitemShut
  {NoStop}%
\bibitem [{\citenamefont {Izrailev}\ \emph
  {et~al.}(2005){\natexlab{b}}\citenamefont {Izrailev}, \citenamefont
  {Makarov},\ and\ \citenamefont {Rend\'on}}]{IzrMkrRnd_PhysRevB_2005}%
  \BibitemOpen
  \bibfield  {author} {\bibinfo {author} {\bibfnamefont {F.~M.}\ \bibnamefont
  {Izrailev}}, \bibinfo {author} {\bibfnamefont {N.~M.}\ \bibnamefont
  {Makarov}}, \ and\ \bibinfo {author} {\bibfnamefont {M.}~\bibnamefont
  {Rend\'on}},\ }\Doi {10.1103/PhysRevB.72.041403} {\bibfield  {journal}
  {\bibinfo  {journal} {Phys. Rev. B},\ }\textbf {\bibinfo {volume} {72}},\
  \bibinfo {pages} {041403(R)} (\bibinfo {year}
  {2005}{\natexlab{b}})}\BibitemShut {NoStop}%
\bibitem [{\citenamefont {Izrailev}\ \emph {et~al.}(2006)\citenamefont
  {Izrailev}, \citenamefont {Makarov},\ and\ \citenamefont
  {Rend\'on}}]{IzrMkrRnd_PhysRevB_2006}%
  \BibitemOpen
  \bibfield  {author} {\bibinfo {author} {\bibfnamefont {F.~M.}\ \bibnamefont
  {Izrailev}}, \bibinfo {author} {\bibfnamefont {N.~M.}\ \bibnamefont
  {Makarov}}, \ and\ \bibinfo {author} {\bibfnamefont {M.}~\bibnamefont
  {Rend\'on}},\ }\Doi {10.1103/PhysRevB.73.155421} {\bibfield  {journal}
  {\bibinfo  {journal} {Phys. Rev. B},\ }\textbf {\bibinfo {volume} {73}},\
  \bibinfo {pages} {155421} (\bibinfo {year} {2006})}\BibitemShut {NoStop}%
\bibitem [{\citenamefont {Rend\'on}\ \emph {et~al.}(2007)\citenamefont
  {Rend\'on}, \citenamefont {Izrailev},\ and\ \citenamefont
  {Makarov}}]{RndIzrMkr_PhysRevB_2007}%
  \BibitemOpen
  \bibfield  {author} {\bibinfo {author} {\bibfnamefont {M.}~\bibnamefont
  {Rend\'on}}, \bibinfo {author} {\bibfnamefont {F.~M.}\ \bibnamefont
  {Izrailev}}, \ and\ \bibinfo {author} {\bibfnamefont {N.~M.}\ \bibnamefont
  {Makarov}},\ }\Doi {10.1103/PhysRevB.75.205404} {\bibfield  {journal}
  {\bibinfo  {journal} {Phys. Rev. B},\ }\textbf {\bibinfo {volume} {75}},\
  \bibinfo {pages} {205404} (\bibinfo {year} {2007})}\BibitemShut {NoStop}%
\bibitem [{\citenamefont {Rend\'on}\ \emph {et~al.}(2011)\citenamefont
  {Rend\'on}, \citenamefont {Makarov},\ and\ \citenamefont
  {Izrailev}}]{RndMkrIzr_PhysRevE_2011}%
  \BibitemOpen
  \bibfield  {author} {\bibinfo {author} {\bibfnamefont {M.}~\bibnamefont
  {Rend\'on}}, \bibinfo {author} {\bibfnamefont {N.~M.}\ \bibnamefont
  {Makarov}}, \ and\ \bibinfo {author} {\bibfnamefont {F.~M.}\ \bibnamefont
  {Izrailev}},\ }\Doi {10.1103/PhysRevE.83.051124} {\bibfield  {journal}
  {\bibinfo  {journal} {Phys. Rev. E},\ }\textbf {\bibinfo {volume} {83}},\
  \bibinfo {pages} {051124} (\bibinfo {year} {2011})}\BibitemShut {NoStop}%
\bibitem [{\citenamefont {Izrailev}\ and\ \citenamefont
  {Makarov}(2001)}]{IzrMkr_OptLett_2001}%
  \BibitemOpen
  \bibfield  {author} {\bibinfo {author} {\bibfnamefont {F.~M.}\ \bibnamefont
  {Izrailev}}\ and\ \bibinfo {author} {\bibfnamefont {N.~M.}\ \bibnamefont
  {Makarov}},\ }\href@noop {} {\bibfield  {journal} {\bibinfo  {journal}
  {Optics Letters},\ }\textbf {\bibinfo {volume} {26}},\ \bibinfo {pages}
  {1604–} (\bibinfo {year} {2001})}\BibitemShut {NoStop}%
\bibitem [{\citenamefont {Izrailev}\ and\ \citenamefont
  {Makarov}(2003)}]{IzrMak_PhysRevB_2003}%
  \BibitemOpen
  \bibfield  {author} {\bibinfo {author} {\bibfnamefont {F.~M.}\ \bibnamefont
  {Izrailev}}\ and\ \bibinfo {author} {\bibfnamefont {N.~M.}\ \bibnamefont
  {Makarov}},\ }\Doi {10.1103/PhysRevB.67.113402} {\bibfield  {journal}
  {\bibinfo  {journal} {Phys. Rev. B},\ }\textbf {\bibinfo {volume} {67}},\
  \bibinfo {pages} {113402} (\bibinfo {year} {2003})}\BibitemShut {NoStop}%
\bibitem [{\citenamefont {Izrailev}\ and\ \citenamefont
  {Makarov}(2005)}]{IzrMkr_JPhysA_2005}%
  \BibitemOpen
  \bibfield  {author} {\bibinfo {author} {\bibfnamefont {F.~M.}\ \bibnamefont
  {Izrailev}}\ and\ \bibinfo {author} {\bibfnamefont {N.~M.}\ \bibnamefont
  {Makarov}},\ }\href@noop {} {\bibfield  {journal} {\bibinfo  {journal} {J.
  Phys. A: Math. Gen.},\ }\textbf {\bibinfo {volume} {38}},\ \bibinfo {pages}
  {10613–} (\bibinfo {year} {2005})}\BibitemShut {NoStop}%
\bibitem [{\citenamefont {Kuhl}\ \emph {et~al.}(2008)\citenamefont {Kuhl},
  \citenamefont {Izrailev},\ and\ \citenamefont
  {Krokhin}}]{KhlIzrKrk_PhysRevLett_2008}%
  \BibitemOpen
  \bibfield  {author} {\bibinfo {author} {\bibfnamefont {U.}~\bibnamefont
  {Kuhl}}, \bibinfo {author} {\bibfnamefont {F.~M.}\ \bibnamefont {Izrailev}},
  \ and\ \bibinfo {author} {\bibfnamefont {A.~A.}\ \bibnamefont {Krokhin}},\
  }\Doi {10.1103/PhysRevLett.100.126402} {\bibfield  {journal} {\bibinfo
  {journal} {Phys. Rev. Lett.},\ }\textbf {\bibinfo {volume} {100}},\ \bibinfo
  {pages} {126402} (\bibinfo {year} {2008})}\BibitemShut {NoStop}%
\bibitem [{\citenamefont {Dietz}\ \emph {et~al.}(2011)\citenamefont {Dietz},
  \citenamefont {Kuhl}, \citenamefont {St\"ockmann}, \citenamefont {Makarov},\
  and\ \citenamefont {Izrailev}}]{DtzKhlStcMkrIzr_PhysRevB_2011}%
  \BibitemOpen
  \bibfield  {author} {\bibinfo {author} {\bibfnamefont {O.}~\bibnamefont
  {Dietz}}, \bibinfo {author} {\bibfnamefont {U.}~\bibnamefont {Kuhl}},
  \bibinfo {author} {\bibfnamefont {H.-J.}\ \bibnamefont {St\"ockmann}},
  \bibinfo {author} {\bibfnamefont {N.~M.}\ \bibnamefont {Makarov}}, \ and\
  \bibinfo {author} {\bibfnamefont {F.~M.}\ \bibnamefont {Izrailev}},\ }\Doi
  {10.1103/PhysRevB.83.134203} {\bibfield  {journal} {\bibinfo  {journal}
  {Phys. Rev. B},\ }\textbf {\bibinfo {volume} {83}},\ \bibinfo {pages}
  {134203} (\bibinfo {year} {2011})}\BibitemShut {NoStop}%
\bibitem [{\citenamefont {Tsang}\ \emph {et~al.}(2001)\citenamefont {Tsang},
  \citenamefont {Kong}, \citenamefont {Ding},\ and\ \citenamefont
  {Ao}}]{TsnKngDng_book_2001}%
  \BibitemOpen
  \bibfield  {author} {\bibinfo {author} {\bibfnamefont {L.}~\bibnamefont
  {Tsang}}, \bibinfo {author} {\bibfnamefont {J.~A.}\ \bibnamefont {Kong}},
  \bibinfo {author} {\bibfnamefont {K.-H.}\ \bibnamefont {Ding}}, \ and\
  \bibinfo {author} {\bibfnamefont {C.~O.}\ \bibnamefont {Ao}},\ }\href@noop {}
  {\emph {\bibinfo {title} {Scattering of electromagnetic waves. Numerical
  simulations}}}\ (\bibinfo  {publisher} {John Wiley and Sons},\ \bibinfo
  {address} {New York},\ \bibinfo {year} {2001})\BibitemShut {NoStop}%
\end{thebibliography}%

\end{document}